\documentclass[a4paper,11pt]{article}

\usepackage{amsmath}
\usepackage{amsfonts}
\usepackage{amssymb}
\usepackage{pstcol}
\usepackage{multido}
\usepackage[all]{xy}

\definecolor{light-blue}{rgb}{0.5,0.9,1}

\newcommand{\Show}[2]{\psframebox[framearc=0.25,fillstyle=solid,fillcolor=#1]{\txt{#2}}}

\newcommand{\ShowXf}[3]{{\Show{#1}{\begin{minipage}[t]{#2}\center{#3}\end{minipage}}}}

\begin{document}

\title{On the interpretation of Michelson--Morley experiments}

\author{Claus L{\"a}mmerzahl$^1$ and Mark P. Haugan$^2$\\
$^1$ Institute for Experimental Physics, Heinrich--Heine--University D\"usseldorf, \\
40225 D\"usseldorf, Germany \\
$^2$ Purdue University, West Lafayette, IN 47907, USA}

\maketitle

\begin{abstract}
Recent proposals for improved optical tests of Special Relativity 
have renewed interest in the interpretation of such tests.  In this 
paper we discuss the interpretation of modern realizations of the 
Michelson--Morley experiment in the context of a new model of 
electrodynamics featuring a vector--valued photon mass.  This model 
is gauge invariant, unlike massive--photon theories based on the Proca 
equation, and it predicts anisotropy of both the speed of light and 
the electric field of a point charge.  The latter leads to an 
orientation dependence of the length of solid bodies which must be 
accounted for when interpreting the results of a Michelson--Morley 
experiment.  Using a simple model of ionic solids we show that, 
in principle, the effect of orientation dependent length can 
conspire to cancel the effect of an anisotropic speed of light 
in a Michelson--Morley experiment, thus, complicating the 
interpretation of the results.  
\end{abstract}

keywords: Michelson--Morley experiment, isotropy of space, 
Maxwell equations, test of Special Relativity


\section{Introduction}

The Michelson--Morley experiment \cite{MichelsonMorley87} is one of 
the most famous physics experiments ever performed, and it has a 
unique place in both the history and the empirical foundations of 
Special Relativity.  Originally conceived as a test for the existence 
of the ether, a medium that was to carry electromagnetic waves and 
whose rest frame would realize Newton's absolute space, the 
experiment's failure to detect the expected anisotropy 
of the speed of light in a frame moving through the ether 
presented physicists with the puzzle eventually solved by 
Einstein's Special Theory of Relativity \cite{Einstein05}.  
Subsequently, Robertson developed a kinematical test theory 
\cite{Robertson49} in which he could show that the 
null results of the Michelson--Morley and Kennedy--Thorndike 
experiments in combination with the result of the time--dilation 
experiment of Ives and Stilwell thoroughly tested the validity 
of Special Relativity.  

The original Michelson--Morley experiment \cite{MichelsonMorley87} 
exploited an interferometer in which a light beam was split into two 
beams propagating back and forth along perpendicular paths of equal 
length.  It was expected that when the interferometer was rotated 
interference of the recombined beams would show the effects of 
an anisotropy of the speed of light caused by the Earth's motion 
through the ether.  The experiment's failure to detect this effect 
had an impact on physicists of the day that is difficult to 
appreciate fully now.  

Modern realizations of the Michelson--Morley experiment 
generally exploit microwave cavities.  For example, Brillet 
and Hall \cite{BrilletHall79} mounted such a cavity on a 
rotating table.  Any orientation dependence of the speed of 
light would lead to a corresponding orientation dependence of 
the frequency of microwaves resonant with the cavity.  The 
null result of the Brillet--Hall experiment, which looked for 
such dependence, constrains any anisotropy of the speed of 
light to be less than a part in $10^{15}$, that is, 
$\delta c/c \lesssim 10^{-15}$.  Proposed space--based versions 
of the Michelson--Morley experiment called 
SUMO \cite{Buchmanetal98} and OPTIS \cite{Laemmerzahletal00}, 
which exploit cryogenic and room--temperature cavity resonators, 
respectively, are capable of tightening this already impressive 
constraint by several orders of magnitude.  The prospect of such 
precise new tests motivates our reexamination of their meaning for 
the empirical foundations of Special Relativity.  

Elementary considerations establish that the angular frequencies 
of electromagnetic waves resonant within a cavity of length 
$L$ are $\omega = c k$, where $k = \pi n/L$, $n = 1, 2, 3, \ldots$, 
denotes the wave number of the resonant wave.  Clearly, any 
orientation dependence of the speed of light, $c = c(\vartheta)$, 
leads to a corresponding orientation dependence of the resonant 
frequencies, 
\begin{equation}
\omega(\vartheta) = c(\vartheta) \frac{n \pi}{ L} \, . \label{Eq:1}
\end{equation}
See Fig. 1 for a simple experimental setup sensitive to 
this sort of dependence.  

From the form of (\ref{Eq:1}) it is clear that an orientation dependence 
of the length of a cavity would also contribute to an orientation 
dependence of resonant frequencies.  In this Letter we show that 
consistent, physically motivated dynamical models can predict both 
an orientation dependence of the length of solid bodies and an 
anisotropy of the speed of light, thus, complicating the 
interpretation of the results of Michelson--Morley experiments.  
For the particular model we analyze below that these 
competing influences on a cavity's resonant frequencies 
can conspire to cancel.  However, we also show that this 
cancellation is negligible in Michelson--Morley experiments 
performed to date or in proposed new versions of the experiment.  
The familiar interpretation of the results of these experiments remains 
valid.  

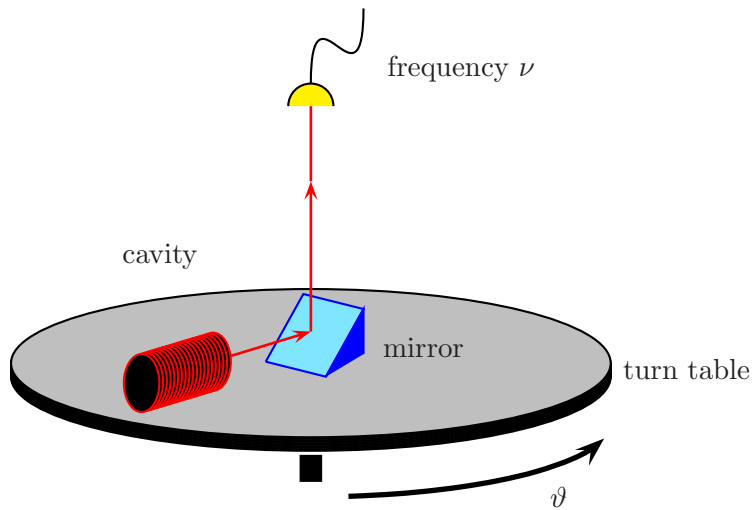
\begin{figure}[t]
\psset{unit=1cm}
\begin{center}
\begin{pspicture}(1,0)(9,7)%\showgrid
\multips(5,1.4)(0,0.02){10}{\psellipse[linecolor=black,fillstyle=solid,fillcolor=lightgray](0,0)(4,1)}
\psline[linecolor=blue,fillstyle=solid,fillcolor=light-blue](4.4,1.6)(5.2,1.4)(5.7,2.3)(4.9,2.5)(4.4,1.6)
\psline[linecolor=blue,fillstyle=solid,fillcolor=blue](5.2,1.4)(5.7,2.3)(5.7,1.7)
\psline[linewidth=1pt,linecolor=red]{->}(2.7,1.3)(5,2)
\psline[linewidth=1pt,linecolor=red]{->}(5,2)(5,4)
\psline[linewidth=1pt,linecolor=red]{-}(5,4)(5,5)
\multips(3.7,1.6)(-0.05,-0.015){20}{\psellipse[linecolor=red,fillstyle=solid,fillcolor=black](0,0)(0.25,0.4)}
\psline[linewidth=0.3cm](5,0)(5,0.36)
\put(5,5){\pswedge[fillstyle=solid,fillcolor=yellow,linestyle=none](0,0){0.3}{0}{180}
\psarc(0,0){0.3}{0}{180}
\psbezier(0,0.3)(0,1.8)(0.7,-0.2)(0.7,1.3)}
\rput(7,5.5){frequency $\nu$}
\rput(6.5,1.8){mirror}
\rput(3,3){\txt{cavity}}
\rput(10,1.5){turn table}
\rput(5,0.8){\pscurve[linewidth=2pt]{->}(0.5,-0.992157)(1,-0.968246)(1.5,-0.927025)(2,-0.866025)(2.5,-0.780625)(3,-0.661438)(3.5,-0.484123)(3.7,-0.379967)(3.9,-0.222205)}
\rput(8.3,-0.2){$\vartheta$}
\end{pspicture}
\end{center}
\caption{A schematic view of Michelson--Morley experiment using a cavity.}
\end{figure}

In the following we describe the dynamical model that is the 
basis for our analysis, show that it predicts both an anisotropic 
speed of light and an orientation dependent length of crystalline 
structures and demonstrate that only when both of these effects 
are accounted for does one obtain a reliable interpretation of 
the results of Michelson-Morley experiments.  

\section{An alternative model of electrodynamics}

Designing and interpreting the results of experimental tests 
of Special Relativity requires that one consider consistent, 
physically reasonable alternatives to familiar relativistic 
dynamics.  In proposing an alternative electrodynamics based 
on modified Maxwell equations we aim to explore the consequences 
of conceivable departures from special relativistic 
physics.  We make no special claim for the correctness 
of any particular alternative that we consider.  We do show 
that these alternatives can lead not only to an anisotropic speed 
of light but also to an anisotropy in the electromagnetic interactions 
that affects the structure of solids.  Only when both of these 
effects are accounted for does one obtain a reliable 
interpretation of the results of Michelson--Morley experiments.

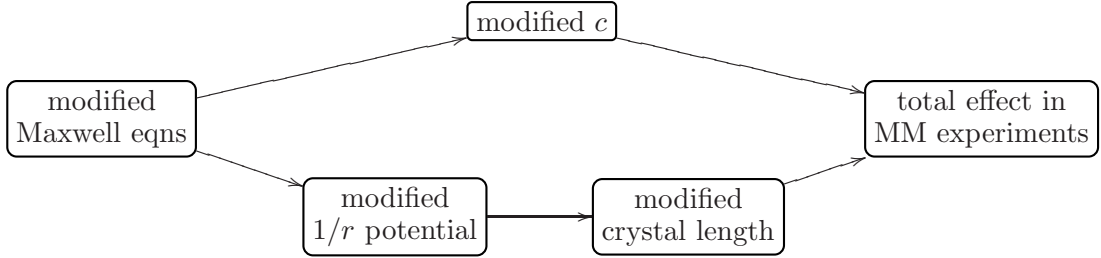
\begin{equation*}
\begin{xy} 0;<1.3cm,0cm>:
,(0,0)*{\ShowXf{white}{3cm}{\txt{modified \\ Maxwell eqns}}}="MM"
,(4.5,1)*{\ShowXf{white}{3cm}{\txt{modified $c$}}}="mc"
,(3,-1)*{\ShowXf{white}{3cm}{\txt{modified \\ $1/r$ potential}}}="mp"
,(6,-1)*{\ShowXf{white}{3cm}{\txt{modified \\ crystal length}}}="ml"
,(9,0)*{\ShowXf{white}{3cm}{\txt{total effect in \\ MM experiments}}}="te"
\ar@{->} "MM";"mc"
\ar@{->} "MM";"mp"
\ar@{->} "mp";"ml"
\ar@{->} "mc";"te"
\ar@{->} "ml";"te"
\end{xy}
\end{equation*}

\subsection{The modified Maxwell equations}

A constructive approach to Maxwell's equations 
\cite{LaemmerzahlHehlPuntigam99,HauganLaemmerzahl00,
HauganLaemmerzahl00a,HauganLaemmerzahl00b} suggests the following 
general structure for equations of motion governing the dynamics 
of the electromagnetic field,  
\begin{equation}
4 \pi j^a = \underbrace{\chi^{abcd} \partial_b 
F_{cd}}_{\hbox{\footnotesize structure of null cones}} + 
\underbrace{\chi^{acd} F_{cd}}_{\hbox{\footnotesize mass--term}} 
\, .\label{GME}
\end{equation}
Here $F_{ab} = \partial_a A_b - \partial_b A_a$ denotes the tensor of 
electromagnetic field strengths with $A_a$ the electromagnetic 
4--potential.  Recent speculations regarding modifications of 
electrodynamics induced by quantum gravity are consistent with this 
general structure (\ref{GME}), see \cite{HauganLaemmerzahl00a} for a review.

The first term on the right--hand side of (\ref{GME}) governs 
the propagation of field discontinuities and thus defines the 
light--cones.  The factor $\chi^{abc}$ in the second term acts 
like a sort of tensorial photon mass.  It gives rise to dispersion 
and to a Yukawa--like form of the field of a point 
charge (see below).  Note that equation (\ref{GME}) is manifestly 
gauge invariant.  In contrast, models of a massive photon based 
on the Proca equation break gauge invariance \cite{ItzyksonZuber80}.

In this Letter we consider the following simple case of 
the equations (\ref{GME}), 
\begin{equation}
4 \pi j^b = \partial_a F^{ab} + \theta_a F^{ab} \, ,\label{model}
\end{equation}
where $\chi^{abcd} = \eta^{a[c} \eta^{d]b}$ and 
$\chi^{acd} = \eta^{ad} \theta^c$ with $\eta^{ab}$ the usual 
Minkowski metric $\eta^{ab} = \hbox{diag(+ -- -- --)}$.  The 
factor $\theta_a$ acts like a sort of vectorial photon mass, though 
it might reflect properties of the vacuum or the value of a 
background field rather than an intrinsic property of the photon 
\cite{HauganLaemmerzahl00b}.  Because of the empirical successes to 
date of the familiar Maxwell's equations, we expect the 
magnitudes of the components $\theta_a$ to be small in a laboratory 
in the Earth's neighborhood.  

A 3+1 decomposition of our modified Maxwell equations (\ref{model}) 
yields 
\begin{eqnarray}
4 \pi \rho & = & \mbox{\boldmath$\nabla$} \cdot \mbox{\boldmath$E$} 
+ \mbox{\boldmath$\theta$} \cdot \mbox{\boldmath$E$} \\
\frac{4 \pi}{c} \mbox{\boldmath$j$} & = & \frac{1}{c} 
\partial_t \mbox{\boldmath$E$} + \mbox{\boldmath$\nabla$} \times 
\mbox{\boldmath$B$} - \frac{1}{c} \theta_0 \mbox{\boldmath$E$} 
+ \mbox{\boldmath$\theta$} \times \mbox{\boldmath$B$} \, .
\end{eqnarray}
where, as usual, $E_i = F_{0i}$ and $B_i = F_{jk}$, for cyclic $i$, 
$j$ and $k$, and where the units of $\theta_0$ and 
$\mbox{\boldmath$\theta$}$ are ${\hbox{sec}}^{-1}$ and 
${\hbox{m}}^{-1}$, respectively.  The physical significance of the 
speed--of--light parameter $c$ is discussed below.  Note that, in 
general, charge is not conserved.  However, this implies no 
inconsistency of our model \cite{HauganLaemmerzahl00b} and 
is, conceivably, a desirable feature in brane worlds where 
matter can tunnel to the bulk \cite{DubovskyRubakovTinyakov00}.

Experimental constraints on a hypothetical photon mass are quite 
severe.  The tightest constraint from a laboratory experiment is 
$m \leq 10^{-47}\;\hbox{g}$, see \cite{WilliamsFallerHill71}, while 
the constraint $m \leq 10^{-48}\;\hbox{g}$ is inferred from measurements 
of the earth's magnetic field.  Recent astrophysical observations 
of electromagnetic radiation of vastly different frequencies emitted by  
GRBs imply the constraint $m \leq 10^{-49}\;\hbox{g}$, see \cite{Schaefer99}.  
We see below that these constraints on $m$ are equivalent to 
constraints on $|\mbox{\boldmath$\theta$}| \hbar / (2 c)$.  

We are interested in laboratory experiments conducted within limited 
regions of spacetime so we will treat $\theta_a$ as constant 
in the following.  In the case of timelike $\theta_a$ there is 
a preferred frame in which only the $\theta_0$ component is nonzero and 
in which electrodynamics is isotropic.  In a frame moving with velocity 
$\mbox{\boldmath$w$}$ relative to this preferred frame 
$\mbox{\boldmath$\theta$}$ is roughly 
$\theta_0 \mbox{\boldmath$w$}$.  This case falls within the range of 
possibilities considered kinematically by Robertson \cite{Robertson49}.  
The case of a spacelike $\theta_a$, in which there is a preferred frame 
in which $\theta_0$ vanishes and electrodynamics is intrinsically 
anisotropic, does not.  

\subsection{Plane electromagnetic waves}

The wave equation that follows from (\ref{model}) is
\begin{equation}
4 \pi \partial_{[a} j_{b]} = \frac{1}{2} \left(\square F_{ba} + 
\theta^c \partial_c F_{ba}\right). \label{WaveEqn}
\end{equation}
For a plane wave, $F_{ab} = F_{ab}^0 e^{- i k_c x^c}$, it implies 
the dispersion relation 
\begin{equation}
\eta^{ab} k_a k_b + i \theta^a k_a = 0 \, .
\end{equation}
Solving for the wave's angular frequency we find 
\begin{equation}
\omega = -\frac{i}{2} \theta^0 \pm c \sqrt{{\mbox{\boldmath$k$}}^2 
- i \theta^{\hat a} k_{\hat a} - \frac{1}{4} (\theta^0)^2} \, .
\end{equation}
which, in the physically interesting limit of small $\theta_a$, is 
approximately 
\begin{equation}
\omega \approx \pm c |\mbox{\boldmath$k$}| \left(1 + \frac{1}{8 
|\mbox{\boldmath$k$}|^2} ( (\mbox{\boldmath$\theta$} \cdot 
\widehat{\mbox{\boldmath$k$}})^2 - c^2 \theta_0^2)\right) 
- \frac{i}{2} \left(\theta^0 \pm c \mbox{\boldmath$\theta$} 
\cdot \widehat{\mbox{\boldmath$k$}}\right) \, . \label{DispersionReal}
\end{equation}
Notice the possibility of weak damping or anti--damping of plane 
waves and of both isotropic and anisotropic dispersion.  

The group velocity corresponding to the dispersion relation 
(\ref{DispersionReal}) is 
\begin{equation}
{\mbox{\boldmath$c$}}_{\rm g} = c 
{\mbox{\boldmath$\nabla$}}_{\mbox{\boldmath$\scriptstyle k$}} 
\hbox{Re}\omega = c \widehat{\mbox{\boldmath$k$}} 
\left( 1 + \frac{1}{8 |\mbox{\boldmath$k$}|^2} 
\left( c^2 \theta_0^2 - 3 (\mbox{\boldmath$\theta$} \cdot 
\widehat{\mbox{\boldmath$k$}})^2 \right) \right) + 
c \frac{\mbox{\boldmath$\theta$} \cdot 
\widehat{\mbox{\boldmath$k$}}}{4 |\mbox{\boldmath$k$}|^2} 
\mbox{\boldmath$\theta$} + {\cal O}(\theta^3/|\mbox{\boldmath$k$}|^3)\, .
\end{equation}
Its modulus is
\begin{equation}
c_{\rm g} = c + \frac{c}{4 |\mbox{\boldmath$k$}|^2} 
\left(c^2 \theta_0^2 - (\mbox{\boldmath$\theta$} \cdot 
\widehat{\mbox{\boldmath$k$}})^2\right)\, .
\end{equation}

The phase velocity implied by (\ref{DispersionReal}) is also anisotropic 
and depends on wavelength, 
\begin{equation}
c_{\rm p} = \frac{\hbox{Re}\omega}{|\mbox{\boldmath$k$}|} 
\approx c + \frac{c}{8 |\mbox{\boldmath$k$}|^2} ( 
(\mbox{\boldmath$\theta$} \cdot 
\widehat{\mbox{\boldmath$k$}})^2 - c^2 \theta_0^2)  \, . \label{phasec}
\end{equation}

In the limit of small wavelengths both the group and phase 
velocities approach $c$ so that this parameter has a clear 
operational meaning.

\subsection{The field of a point charge}

For a point charge $q$ at rest $\mbox{\boldmath$j$} = 0$ and 
the modified Maxwell equations (4) and (5) take the form 
\begin{eqnarray}
4 \pi q \delta(\mbox{\boldmath$x$}) & = & \mbox{\boldmath$\nabla$} 
\cdot \mbox{\boldmath$E$} + \mbox{\boldmath$\theta$} 
\cdot \mbox{\boldmath$E$} \label{PointCharge1} \\
0 & = & \mbox{\boldmath$\nabla$} \times \mbox{\boldmath$B$} 
+ \frac{1}{c} \partial_t \mbox{\boldmath$E$} - 
\frac{1}{c} \theta_0 \mbox{\boldmath$E$} + 
\mbox{\boldmath$\theta$} \times \mbox{\boldmath$B$} \, . \label{PointCharge2}
\end{eqnarray}
Interestingly, when $\theta_0$ is nonzero there is no static solution 
to these equations because of charge nonconservation.  In that case, 
$\frac{d}{dt} q = \theta_0 q$ which implies a corresponding time 
dependence of the electric field, 
$\frac{d}{dt} \mbox{\boldmath$E$} = \theta_0 \mbox{\boldmath$E$}$.  
Consequently, the terms in (\ref{PointCharge2}) involving the electric 
field cancel, so that $\mbox{\boldmath$B$} = 0$.  We can then solve 
equation (\ref{PointCharge1}) by introducing a scalar potential, 
$\mbox{\boldmath$E$} = - \mbox{\boldmath$\nabla$} \phi$, to obtain 
\begin{equation}
\Delta\phi + \mbox{\boldmath$\theta$} \cdot \mbox{\boldmath$\nabla$} 
\phi = 4 \pi q \delta(\mbox{\boldmath$x$}) \, . \label{EquationForPhi}
\end{equation}
The solution of this equation is \cite{HauganLaemmerzahl00b}
\begin{equation}
\phi(\mbox{\boldmath$x$}) = - q \frac{e^{- \frac{1}{2} 
(\mbox{\boldmath$\scriptstyle \theta$} \cdot 
\mbox{\boldmath$\scriptstyle x$} + 
|\mbox{\boldmath$\scriptstyle \theta$}| r)}}{r} = 
- q \frac{e^{- \frac{1}{2} 
|\mbox{\boldmath$\scriptstyle \theta$}| r 
(1 + \cos\vartheta)}}{r} \, , \label{Eq:EDyn:AnisMassPotential}
\end{equation}
where $r = |\mbox{\boldmath$x$}|$ and $\vartheta$ is the angle 
between $\mbox{\boldmath$\theta$}$ and the position vector 
$\mbox{\boldmath$x$}$.  Note the anisotropic, Yukawa--like 
character of this potential when $|\mbox{\boldmath$\theta$}| \neq 0$.  

The corresponding electric field is 
\begin{equation}
\mbox{\boldmath$E$}(\mbox{\boldmath$x$}) = - q 
\mbox{\boldmath$\nabla$} \cdot \phi(\mbox{\boldmath$x$}) = 
- q e^{- \frac{1}{2} (\mbox{\boldmath$\scriptstyle \theta$} 
\cdot \mbox{\boldmath$\scriptstyle x$} + 
|\mbox{\boldmath$\scriptstyle \theta$}| r)} 
\left(\frac{\mbox{\boldmath$x$}}{r^3} + \frac{1}{2} 
\left(\mbox{\boldmath$\theta$} + |\mbox{\boldmath$\theta$}| 
\frac{\mbox{\boldmath$x$}}{r}\right) \frac{1}{r}\right) 
\, . \label{SolElectricField}
\end{equation}

\section{The structure of ionic solids}

In Michelson--Morley experiments light propagates back and forth 
along interferometer arms or back and forth within a resonant 
cavity.  Since such structures are made of solids whose 
properties reflect the electromagnetic interaction between atoms 
we must consider the effect our modification of the Maxwell equations 
has on the structure of Michelson--Morley experimental apparatus.  In 
this Letter we do so by considering the illustrative case of ionic 
crystalline solids.  In this section we show that the our 
modified Maxwell equations can predict an orientation dependence 
of the separation between crystal planes and, thus, of the length 
of ionic solids.  

As a simple model of an ionic crystal we take an infinite chain of 
alternating positive and negative ions with charges of magnitude $q$.  
To calculate this crystal's structure we proceed as in 
\cite{Kittel71}.  Our scalar potential 
(\ref{Eq:EDyn:AnisMassPotential}) governs the electric 
interaction between ions.  To stabilize the crystal there must 
also be a repulsive potential which we model by $\lambda 
e^{- r/\rho}$ where $\rho > 0$ and $\lambda > 0$ are parameters 
characterizing, respectively, the potential's range and strength.  
Fundamentally, this stabilizing potential is 
a reflection of the Pauli exclusion principle \cite{Kittel71}.  
While one could consider the effect of conceivable modifications 
of quantum physics on the outcomes of Michelson--Morley experiments, 
here we consider only effects caused by our modifications of 
the Maxwell equations.  Therefore, we take the parameters 
$\rho $ and $\lambda $ to be independent of $\theta_a$.  

The total energy for an ion in our infinite model crystal is
\begin{equation}
U = \sum_{n = 1}^\infty \left(\lambda e^{- n r/\rho} + 
(-1)^n q^2 \frac{e^{- \frac{1}{2} 
|\mbox{\boldmath$\scriptstyle \theta$}| n r 
(1 + \cos\vartheta)}}{n r}\right) + \sum_{n = 1}^\infty 
\left(\lambda e^{- n r/\rho} + (-1)^n q^2 \frac{e^{- 
\frac{1}{2} |\mbox{\boldmath$\scriptstyle \theta$}| n r 
(1 - \cos\vartheta)}}{n r}\right) \, ,
\end{equation}
where $\vartheta$ denotes the angle between our ionic chain and 
the vector field $\mbox{\boldmath$\theta$}$.  The sum is easily 
computed, 
\begin{equation}
U = \frac{2 \lambda}{1 - e^{- r/\rho}} - \frac{q^2}{r} 
\ln\left(1 + e^{- \frac{1}{2} 
|\mbox{\boldmath$\scriptstyle \theta$}| r 
(1 + \cos\vartheta)}\right) - \frac{q^2}{r} 
\ln\left(1 + e^{- \frac{1}{2} |\mbox{\boldmath$\scriptstyle \theta$}| 
r (1 - \cos\vartheta)}\right) \, .
\end{equation}
The logarithms yield the Madelung constant which, in this case, depends 
on $\mbox{\boldmath$\theta$}$.  The energy expression above reduces 
to the familiar result in the case $\mbox{\boldmath$\theta$} = 0$.  

The equilibrium condition that determines the separation, $R$, between 
ions (crystal planes) is 
\begin{equation}
0 = \left.\frac{d U(r)}{d r}\right|_{R} \, .
\end{equation}
This is an implicit equation for $R$,
\begin{eqnarray}
0 & = & - \frac{\lambda}{\rho} \frac{2 e^{- R/\rho}}{\left(1 - 
e^{- R/\rho}\right)^2} + \frac{q^2}{R^2} \ln\left(1 + 
e^{- \frac{1}{2} |\mbox{\boldmath$\scriptstyle \theta$}| R 
(1 + \cos\vartheta)}\right) + \frac{1}{2} \frac{q^2}{R} 
\frac{|\mbox{\boldmath$\scriptstyle \theta$}| (1 + 
\cos\vartheta) e^{- \frac{1}{2} |\mbox{\boldmath$\scriptstyle \theta$}| 
R (1 + \cos\vartheta)}}{1 + e^{- \frac{1}{2} 
|\mbox{\boldmath$\scriptstyle \theta$}| R (1 + \cos\vartheta)}} 
\nonumber\\
& & + \frac{q^2}{R^2} \ln\left(1 + e^{- \frac{1}{2} 
|\mbox{\boldmath$\scriptstyle \theta$}| R (1 - \cos\vartheta)}\right) 
+ \frac{1}{2} \frac{q^2}{R} \frac{|\mbox{\boldmath$\scriptstyle \theta$}| 
(1 - \cos\vartheta) e^{- \frac{1}{2} |\mbox{\boldmath$\scriptstyle \theta$}| 
R (1 - \cos\vartheta)}}{1 + e^{- \frac{1}{2} 
|\mbox{\boldmath$\scriptstyle \theta$}| R (1 - \cos\vartheta)}} \, .
\end{eqnarray}
which we can simplify because we are interested in the case of small 
$|\mbox{\boldmath$\theta$}| R$ and because $R/\rho$ is large, implying 
$\frac{1}{\rho} \frac{e^{- R/\rho}}{\left(1 
- e^{- R/\rho}\right)^2} \approx  \frac{1}{\rho} e^{- R/\rho}$.  
To leading order in $|\mbox{\boldmath$\theta$}| R$ the simplified 
equation is 
\begin{equation}
R^2 \left(e^{- R/\rho} + \frac{q^2}{32} \frac{\rho}{\lambda} 
|\mbox{\boldmath$\theta$}|^2 (1 + \cos^2\vartheta)\right) =  
q^2 \frac{\rho}{\lambda} \ln 2 \, .
\end{equation}

To determine the fundamental scale $R$ it is convenient to introduce 
$R_0$, the fundamental scale in the case $\mbox{\boldmath$\theta$} = 0$, 
\begin{equation}
R_0^2 e^{- R_0/\rho} = q^2 \frac{\rho}{\lambda} \ln 2 \, . \label{R0}
\end{equation}
Then, $R = R_0 + \delta R$ and solving for $\delta R$ to leading 
order we obtain 
\begin{equation}
\frac{\delta R}{R_0} = \frac{1}{32} \frac{\rho R_0 
|\mbox{\boldmath$\theta$}|^2}{\ln 2 \left(1 - 2 \rho/R_0\right)}  
(1 + \cos^2\vartheta) \, ,
\end{equation}
where the charge $q$ has been eliminated by using (\ref{R0}).

Clearly, when $|\mbox{\boldmath$\theta$}|$ is nonzero the fundamental 
crystal scale $R = R_0 + \delta R$ is orientation dependent.  The 
overall length $L$ of a solid with $n$ crystal planes is 
\begin{equation}
L(\vartheta) = n R_0 \left(1 + \frac{\delta R}{R_0}\right) = 
L_0 \left(1 + \frac{1}{32} \frac{\rho R_0 
|\mbox{\boldmath$\theta$}|^2}{\ln 2 \left(1 - 2 \rho/R_0\right)} 
(1 + \cos^2\vartheta)\right) \, , \label{Ltheta}
\end{equation}
where $L = n R$ and $L_0 = n R_0$.  

Note, in passing, that such orientation dependence of the length of 
solids implies that the spatial coordinates we have been using 
do not have the physical meaning one might have expected.  This is 
of no real concern since we will computing physical observables 
that are, in effect, the result of a direct comparison of a body's 
extension to an electromagnetic wavelength.  Note, too, that one 
might have chosen to consider the Bohr radius of a hydrogen atom 
as a fundamental standard of length.  This has also been shown to 
depend on parameters of test theories encompassing modified 
Maxwell equations \cite{Will74}.  

\section{Predicting the outcome of a Michelson--Morley experiment}

\subsection{Interferometer experiments}

The phase difference monitored in an interferometric Michelson--Morley 
experiment is $\phi = \omega \; \Delta t$, where $\Delta t$ denotes 
the difference in time for propagation at the phase velocity along the 
interferometer's two arms.  This is influenced by any anisotropy of 
the phase velocity of light as well as by any orientation dependence 
of the lengths of the interferometer's arms.  

Our Eq.(\ref{phasec}) gives the phase velocity as
\begin{equation}
c_{\rm p}(\vartheta) = c + \frac{c}{8 |\mbox{\boldmath$k$}|^2} 
(|\mbox{\boldmath$\theta$}|^2 \cos^2\vartheta - c^2 \theta_0^2)
\end{equation}
where $\vartheta$ is the angle between the interferometer arm the 
wave is propagating along and $\mbox{\boldmath$\theta$}$.  
Our Eq.(\ref{Ltheta}) gives the length of each of the 
interferometer's arms as a function of its angle relative to 
$\mbox{\boldmath$\theta$}$.  Consequently, the observed phase 
shift as a function of the angle $\vartheta$ of a chosen arm is 
\begin{eqnarray}
\Delta t & = & \frac{L(\vartheta)}{c_{\rm p}(\vartheta)} - 
\frac{L(\vartheta + \frac{\pi}{2})}{c_{\rm p}(\vartheta + 
\frac{\pi}{2})} \nonumber\\
& = & \frac{L_0 |\mbox{\boldmath$\theta$}|^2}{8 c} 
\left(\frac{1}{4} \frac{\rho R_0}{\ln 2 \left(1 - 2 \rho/R_0\right)} 
- \frac{c^2}{\omega^2}\right) (\cos^2\vartheta - \sin^2\vartheta)  
+ {\cal O}(|\mbox{\boldmath$\theta$}|^4) \, .
\end{eqnarray}

Notice that even when $|\mbox{\boldmath$\theta$}| \neq 0$ it is 
conceivable that there will be {\it no} interference signal.  It 
is possible for the effects of the anisotropic phase velocity 
and the orientation dependence of the length of the interferometer's 
arms to cancel at a particular frequency, namely, the one that satisfies 
\begin{equation}
\frac{1}{\omega^2} - \frac{1}{4 c^2} \frac{\rho R_0}{\ln 2 
\left(1 - 2 \rho/R_0\right)} = 0\, . \label{condInterf}
\end{equation}

Typical crystals have $\rho \sim 0.1 \; R_0$ and 
$R_0 \sim 10^{-10}\;\hbox{m}$ implying a cancellation frequency 
of 
\begin{equation}
\omega = 2 c \sqrt{\frac{0.8 \ln 2}{0.1\; R_0^2}} = 
1.4 \times 10^{19}\;\hbox{Hz}  \, .
\end{equation}
Since interferometric Michelson--Morley experiments are performed 
with visible light having frequencies much lower than this the 
cancellation of competing effects is not a problem in practice.  

\subsection{Resonant cavity experiments}

The frequency monitored in a resonant--cavity Michelson--Morley 
experiment is influenced by any anisotropy of the phase 
velocity of light as well as by any orientation dependence of 
the cavity's length.  The phase velocity governs the relationship 
between the resonant radiation's frequency and wavelength while 
the cavity's length determines the resonant wavelength.  

For a standing wave in a cavity we have
\begin{equation}
|\mbox{\boldmath$k$}(\vartheta)| = 
\frac{\pi n}{L(\vartheta)}\, ,
\end{equation}
where $n = 1, 2, 3, \ldots$ is the number of half wavelengths 
in the standing wave.  The corresponding resonant 
angular frequencies are  
\begin{eqnarray}
\omega(\vartheta) & = & |\mbox{\boldmath$k$}(\vartheta)| c 
\left(1 + \frac{1}{8 |\mbox{\boldmath$k$}|^2} 
(|\mbox{\boldmath$\theta$}|^2 \cos^2\vartheta - 
\theta_0^2)\right) \nonumber\\
& = & \frac{\pi n c}{L_0} \left(1 - \frac{L_0^2}{32 \pi^2 n^2} 
\theta_0^2 - \frac{1}{32} \frac{\rho R_0 
|\mbox{\boldmath$\theta$}|^2}{\ln 2 \left(1 - 2 \rho/R_0\right)} 
\right. \nonumber\\
& & \left. \qquad\qquad + \frac{|\mbox{\boldmath$\theta$}|^2}{32} 
\left(\frac{L_0^2}{\pi^2 n^2} - \frac{\rho R_0}{\ln 2 \left(1 - 
2 \rho/R_0\right)}\right) \cos^2\vartheta\right) + 
{\cal O}(|\mbox{\boldmath$\theta$}|^4)\, .
\end{eqnarray}
The amplitude of the fractional variation in resonant frequency 
as the cavity rotates relative to $\mbox{\boldmath$\theta$}$ is, 
therefore,  
\begin{equation}
\frac{\delta\omega}{\omega} = \frac{|\mbox{\boldmath$\theta$}|^2}{16} 
\left(\frac{L_0^2}{\pi^2 n^2} - \frac{\rho R_0}{\ln 2 \left(1 - 
2 \rho/R_0\right)}\right) \, .  \label{CavVariation}
\end{equation}

Once again, even when $|\mbox{\boldmath$\theta$}| \neq 0$ it is 
conceivable that the effects of an anisotropic speed of light and 
of an orientation dependence of the length of solid bodies can cancel.  
In this case the condition for cancellation is 
\begin{equation}
\frac{L_0^2}{\pi^2 n^2} = \frac{\rho R_0}{\ln 2 \left(1 - 
2 \rho/R_0\right)} \approx 1.8 \times 10^{-21}\;\hbox{$m^2$} \, .
\end{equation}

For a mean cavity length of, say, $L_0 \sim 30\;\hbox{cm}$ this 
implies $n \approx  2.3 \times 10^{9}$.  The corresponding mean 
resonant frequency is 
$\nu_0 = c n/(2 L_0) \sim 1.1 \times 10^{18}\;\hbox{Hz}$.  Since 
resonant cavity Michelson--Morley experiments are performed 
with microwaves having frequencies much lower than this the 
cancellation of competing effects is not a problem in practice.  

\section{Conclusions and outlook}

The design and the interpretation of the results of experimental 
tests of Special Relativity demand consideration of consistent, 
physically reasonable alternatives to familiar relativistic 
dynamics.  One can only design experiments to search for 
differences between the predictions of special relativistic 
and alternative dynamics when one has determined the 
predictions of both.  

In this Letter we have shown that the alternative model of 
electrodynamics based on the modified Maxwell equations (3) 
can predict both an anisotropic speed of light and an 
orientation dependence of the length of solid bodies.  This 
combination of predictions complicates the interpretation of 
interferometric and resonant--cavity Michelson--Morley experiments 
because, in principle, the effect of orientation dependence of 
the lengths of interferometer arms and cavity resonators can 
cancel the effect of an anisotropic speed of light.  Both 
effects must be accounted for to obtain a reliable interpretation 
of the results of a Michelson--Morley experiment.  

We have used the simplest possible model of ionic crystals to 
estimate the orientation dependence of the lengths of solid bodies 
predicted by the modified electrodynamics (3).  While estimates 
based on more sophisticated models of other types of solids need 
to be made, our estimates suggest that for this modified dynamics 
the effect of an anisotropic speed of light dominates that of 
orientation dependence of the length of solid bodies in past 
and proposed Michelson--Morley experiments.  For 
example, the precise microwave experiment of Brillet and 
Hall \cite{BrilletHall79} employed a glass--ceramic cavity 0.305 m long 
operated at a resonant frequency of roughly $10^{14}$ Hz.  Taking the 
estimates of the previous section as an indication of the 
order of magnitude of any predicted orientation dependence of the 
length of the glass--ceramic cavity, even though it is not an ionic 
solid body, we see that the operating frequency is four orders of 
magnitude below that for which the effect of orientation 
dependent cavity length would be large enough to compensate for 
the effect of the predicted anisotropic speed of light.  Similar 
considerations establish that the effect of orientation dependent 
cavity length is negligible in the proposed 
SUMO \cite{Buchmanetal98} and OPTIS \cite{Laemmerzahletal00} 
experiments.  We note that the prediction (\ref{CavVariation}) 
and the target precision of these proposed experiments suggests 
that they will be able to impose the constraint 
$|\mbox{\boldmath$\theta$}| \lesssim  0.3\; {\hbox{m}}^{-1}$.  
Note that is not competitive with astrophysical constaints  
\cite{HauganLaemmerzahl00b}.

We will discuss other predictions of the alternative electrodynamics 
(3) elsewhere \cite{HauganLaemmerzahl00b}.  Similar analyses can be 
performed for the more general model (2) which can predict 
birefringence and can be expected to predict more complicated orientation 
dependence of the speed of light and the lengths of solid bodies.  

One final remark, the results of Hughes--Drever experiments like 
\cite{Chuppetal89} provide some justification for our 
assumption in section 3 that quantum physics remains intrinsically 
isotropic in the alternative models of dynamics we have considered.  
Such experiments can, for example, be shown to tightly 
constrain any anisotropy of the inertial masses of particles 
that appear in the Schr\"odinger equation \cite{Laemmerzahl98}.

\section*{Acknowledgement}

C.L. thanks Ken Nordtvedt, Gerhard Sch\"afer, and Stephan Schiller for 
very useful discussions and the Optikzentrum of the University of 
Konstanz, where part of this work was written, for financial support.


\begin{thebibliography}{10}

\bibitem{MichelsonMorley87}
A.A. Michelson and E.W. Morley.
\newblock On the relative motion of the {E}arth and the luminiferous ether.
\newblock {\em Am.\ J.\ Sci.}, 34:333, 1887.

\bibitem{Einstein05}
A.~Einstein.
\newblock Zur {E}lektrodynamik bewegter {K}{\"o}rper.
\newblock {\em Ann.\ Physik}, 17:891, 1905.

\bibitem{Robertson49}
H.P. Robertson.
\newblock Postulate versus observation in the {S}pecial {T}heory of
  {R}elativity.
\newblock {\em Rev.\ Mod.\ Phys.}, 21:378, 1949.

\bibitem{BrilletHall79}
A.~Brillet and J.L. Hall.
\newblock Improved laser test of the isotropy of space.
\newblock {\em Phys.\ Rev.\ Lett.}, 42:549, 1979.

\bibitem{Buchmanetal98}
S.~Buchman, J.P. Turneaure, J.A. Lipa, M.~Dong, K.M. Cumbernack, and Wang. S.
\newblock A superconducting microwave oscillator clock for use on the space
  station.
\newblock {\em Proceedings of the IEEE International Frequency Symposium},
  IEEE:534, 1998.

\bibitem{Laemmerzahletal00}
C.~L{\"a}mmerzahl, S.~Schiller, H.-J. Dittus, and A.~Peters.
\newblock {OPTIS} -- satellite based optical tests of special and general
  relativity.
\newblock University of D{\"u}sseldorf.

\bibitem{LaemmerzahlHehlPuntigam99}
C.~L{\"a}mmerzahl, R.~Puntigan, and F.W. Hehl.
\newblock Can the electromagnetic field couple to post--{R}iemannian
  structures?
\newblock In T.~Piran and R.~Ruffuni, editors, {\em Proceedings of the Eight
  Marcel Grossmann Meeting on General Relativity}, page 457. World Scientific,
  Singapore, 1999.

\bibitem{HauganLaemmerzahl00}
M.P. Haugan and C.~L{\"a}mmerzahl.
\newblock On the experimental foundations of the {M}axwell equations.
\newblock {\em Ann.\ Phys. (Leipzig)}, 9:119, 2000.

\bibitem{HauganLaemmerzahl00a}
M.P. Haugan and L{\"a}mmerzahl.
\newblock Principles of equivalence: Their role in gravitation physics and
  experiments that test them.
\newblock In C.~L{\"a}mmerzahl, C.W.F. Everitt, and F.W. Hehl, editors, {\em
  Gyros, Clocks, and Interferometers: Testing Relativistic Gravity in Space},
  page to appear. Springer--Verlag, Berlin, 2000.

\bibitem{HauganLaemmerzahl00b}
M.~Haugan and C.~L{\"a}mmerzahl.
\newblock On the mass of the photon.
\newblock in preparation.

\bibitem{ItzyksonZuber80}
C.~Itzykson and J.-B. Zuber.
\newblock {\em Quantum Field Theory}.
\newblock McGraw--Hill, New York, 1980.

\bibitem{DubovskyRubakovTinyakov00}
S.L. Dubovsky, V.A. Rubakov, and P.G. Tinyakov.
\newblock Is the electric charge conserved in brane world?
\newblock gr--qc/0007179.

\bibitem{WilliamsFallerHill71}
E.R. Williams, J.E. Faller, and H.A. Hill.
\newblock New experimental test of {C}oulomb's law: A laboratory upper limit on
  the photon rest mass.
\newblock {\em Phys.\ Rev.\ Lett.}, 26:721, 1971.

\bibitem{Schaefer99}
B.E. Schaefer.
\newblock Severe limits on variations of the speed of light with frequency.
\newblock {\em Phys.\ Rev.\ Lett.}, 82:4964, 1999.

\bibitem{Kittel71}
Ch. Kittel.
\newblock {\em Introduction to solid state physics}.
\newblock Wiley, New York, 1971.

\bibitem{Will74}
C.M. Will.
\newblock Gravitational red--shift measurements as tests of nonmetric theories
  of gravity.
\newblock {\em Phys.\ Rev.}, D 10:2330, 1974.

\bibitem{Chuppetal89}
T.E. Chupp, R.J. Hoara, R.A. Loveman, E.R. Oteiza, J.M. Richardson, and M.E.
  Wagshul.
\newblock Results of a new test of local {L}orentz invariance: A search for
  mass anisotropy in ${}^{21}\hbox{Ne}$.
\newblock {\em Phys.\ Rev.\ Lett.}, 63:1541, 1989.

\bibitem{Laemmerzahl98}
C.~L{\"a}mmerzahl.
\newblock Quantum tests of foundations of general relativity.
\newblock {\em Class.\ Quantum Grav.}, 14:13, 1998.

\end{thebibliography}
\end{document}